\documentclass[dvips]{article}
\usepackage{icrctc07}

\newcommand{\refeq}[1]{Eq.~(\ref{eq:#1})}
\newcommand{\reffig}[1]{Fig.~\ref{fig:#1}}

\newcommand{\Deg}{^{\circ}}
\newcommand{\Sem}{S_{\rm em}}
\newcommand{\gcmsq}{\mbox{g}/\mbox{cm}^2}

\newcommand{\Smu}{S_{\mu}}
\newcommand{\Xmax}{X_{\rm max}}
\newcommand{\Xmaxavg}{\langle X_{\rm max} \rangle}

\newcommand{\Nmu}{N_{\mu}}

\title{Applying Extensive Air Shower Universality to Ground Detector Data}
\shorttitle{EAS Universality}

\authors{Fabian~Schmidt$^{1}$, Maximo~Ave$^{1}$, Lorenzo~Cazon$^{1}$, Aaron~Chou$^{2}$}
\shortauthors{F.~Schmidt and et al.}
\afiliations{$^1$Department of Astronomy and Astrophysics, The University of Chicago, Chicago, IL~~60637\\ $^2$Fermi National Accelerator Laboratory, Batavia, IL~~60510}
\email{fabians@oddjob.uchicago.edu}

\abstract{Air shower universality states that the electromagnetic part of 
hadron-induced
extensive air showers (EAS) can be completely described in terms of the primary
energy and shower age.
In addition, simulations show that the muon part is well characterized by
an overall normalization which depends on the primary particle and hadronic
interaction model. We investigate the consequences of EAS universality
for ground arrays, which sample EAS at large core distances, and show
how universality can be used to experimentally determine the muon
content as well as the primary energy of cosmic ray air showers in a
model-independent way.}

\begin{document}
\maketitle

%\section{Introduction}
%\label{sec:intro}

A ground array detector samples the particles in an Extensive Air Shower
(EAS) at a limited number of points at different distances $r$ from the shower
axis. From this sample, an observable has to be defined to estimate the
shower size. To avoid the large fluctuations in the signal integrated
over all distances caused by fluctuations in the shower development,
Hillas \cite{Hillas} proposed to use the signal at a given distance,
$S(r)$, to determine the shower size.  The optimal distance $r_{\rm opt}$ 
\cite{Newton} where experimental uncertainties in the signal determination 
are minimized is mainly
determined by the experiment geometry (spacing between ground array
detectors). In this paper, we consider the signal in water Cherenkov
detectors as employed by the Auger Observatory ($r_{\rm opt} = 1000$~m).
Similar calculations can be done for any ground array detectors.

Using Monte Carlo simulations, $S(r_{\rm opt})$ is related with the
energy of the incoming cosmic ray. This {\it calibration} suffers from large 
systematics due to uncertainties in the hadronic models and the assumptions 
that have to be made about the primary cosmic ray composition. In this work, 
we propose a new method to determine the {\it calibration} in a model 
independent way. Furthermore, this method allows us to determine the number 
of muons produced in air showers.

The method is based on what we will call {\it air shower universality} \cite{Chou}:
to a remarkable degree of precision, EAS can be characterized by only three 
parameters: the primary energy
$E_0$, the depth of shower maximum $\Xmax$, and the overall
normalization of the muon component $\Nmu$. The parameters
$\Xmax$ and $\Nmu$ are linked to the mass of the primary particle,
ranging from proton to iron, and are subject to significant
shower-to-shower fluctuations. All composition and model dependence is
distilled in these two parameters with clear physical
interpretation. In addition to determining a model-indepedent 
energy estimator, they can be compared with simulations
to infer the cosmic ray composition and place constraints on hadronic
interaction models.

%Air shower universality is a consequence of the huge amount of
%particles involved in an extensive air shower at ultra-high energies,
%which leads to a statistical smoothing of the shower properties after
%the first few interaction lengths. The
%early stages of the shower development only determine the position of the
%shower maximum and the overall muon normalization. 
Previous studies have demonstrated that the energy spectra and angular
 distributions of electromagnetic particles \cite{Nerling,Giller}, as
 well as the lateral distribution of energy deposit close to the
 shower core \cite{Gora} are all universal, i.e. they are functions of
 $E_0$, $\Xmax$, and the atmospheric depth $X$ only (the
 dependence on $X$ and $\Xmax$ is commonly put in terms of the shower
 age $s$).

By exploiting shower universality, we show that it is possible to
separate the known shower properties, including the electromagnetic
particle flux on ground and the average depth of shower maximum $\Xmaxavg$, from the
unknown, the surface detector energy scale and the normalization of
the muon signal at $r_{\rm opt}$ which is tightly
correlated with the overall number of muons in a shower. $\Xmaxavg$ as
a function of energy has been measured with good precision by
fluorescence detectors, and can also be inferred from surface detector
variables.

%This paper is organized as follows: in section 2 we test {\it air shower universality}
% at large distances from the shower axis, in section 3 we describe the method to obtain the surface
% detector energy scale and the number of muons, and in section 4 we discuss the
% systematics.

\section{EAS Universality at large core distances}
\label{sec:univ}

\begin{figure}[t]
\begin{center}
\noindent
\includegraphics [width=0.45\textwidth]{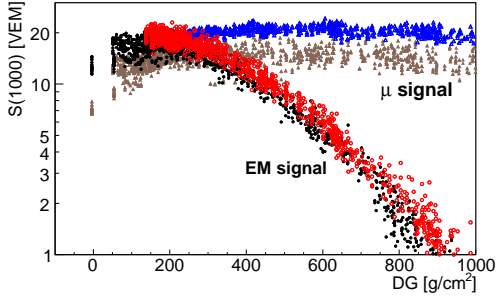}
\end{center}
\caption{Electromagnetic (circles) and muon signals (triangles) at $r=1000$~m in the shower plane vs. distance to ground of the shower maximum, for proton (filled symbols) and iron showers (open symbols) simulated with QGSJetII/Fluka.}
\label{fig:Sempure-DG}
\end{figure}

In this section we will test shower universality in terms of the shower
 plane signal, i.e. the signal generated by particles in a fiducial
flat detector parallel to the shower plane (orthogonal to the shower
axis). By avoiding geometric projection effects, this allows us to 
compare showers at different zenith angles. We have assumed a cylindrical
detector with a top area of 10 m$^2$ and 1.2 m height (similar to the ones
used in the Pierre Auger Observatory). The response of the detector,
simulated using Geant 4, is expressed in units of VEM (the signal of a vertical, central muon).

We have generated a library of showers that span a zenith angle range
of 0$\Deg$ to 70$\Deg$ and an energy range of 10$^{17}$~eV to
10$^{20}$~eV. Showers of proton and iron primaries were generated using
CORSIKA 6.500/6.502 \cite{corsika} and the hadronic interaction models QGSJetII-03 \cite{qgsjet2} and
Fluka \cite{fluka}. In addition, we simulated proton/iron showers at 10$^{19}$ eV 
and different zenith angles using other hadronic interaction models
(QGSJetII-03/Gheisha2002 \cite{gheisha} and Sibyll~2.1/Fluka \cite{sibyll,sibyll2}).

The shower-plane signals were separated into signals from
electromagnetic particles and muons.  We include the signal from the 
electromagnetic decay products of muons ($\sim15$\% of the muon signal)
in the muon component, 
the remaining signal being the `pure' electromagnetic component 
$\Sem$.

\begin{figure}[t]
\begin{center}
\noindent
\includegraphics [width=0.45\textwidth]{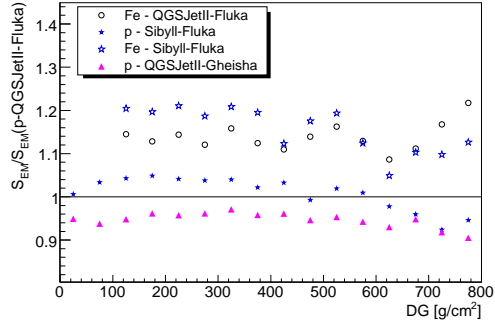}
\end{center}
\caption{Electromagnetic signals (in the shower plane, $r=1000$~m) relative to that
of proton-QGSJetII.}\label{fig:Sempure-ratios}
\end{figure}

\reffig{Sempure-DG} shows the electromagnetic signal for a core distance 
of 1000~m (circles, proton and iron showers) as a function of 
$DG = X_{\rm ground}-\Xmax$,
the distance from the shower maximum to the detector measured along the shower
axis (in $\gcmsq$). Note that this plot contains showers from all zenith 
angles. Apparently, the signals
from proton and iron are very similar, though there is a slight shift
in the overall normalization. This is in violation of shower universality,
which states that showers of the same energy at the same evolutionary
stage (given by $DG$) should have the same electromagnetic component.

\reffig{Sempure-ratios} shows the electromagnetic signal for different
models and primaries, relative to a reference (proton QGSJetII/Fluka). 
Note that the different model predictions for a given primary
are within 5\% of each other. There is, however, a systematic offset of
about 13\% between proton and iron signals. We also found that the systematic
differences in the {\it number density} of particles are smaller, about
8\%. This effect persists also when comparing signals at the same 
shower age instead of $DG$.

\begin{figure}[t]
\begin{center}
\noindent
\includegraphics [width=0.45\textwidth]{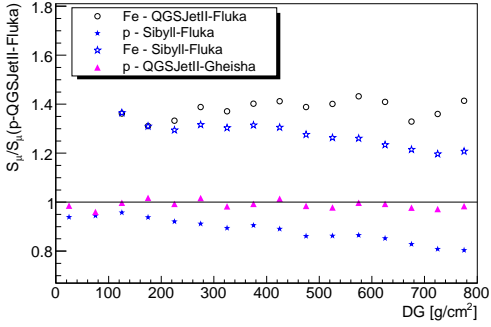}
\end{center}
\caption{Muon ground signals (in the shower plane, $r=1000$~m) relative to that
of proton-QGSJetII vs. distance to ground for different primaries and hadronic
interaction models.}\label{fig:Smu-ratios}
\end{figure}

\reffig{Sempure-DG} also shows the muon signal ($S_{\mu}$, triangles) as a 
function of $DG$ for the same proton and iron showers. 
The dependence of the signal on the 
primary mass ($\sim 40$\% between proton and iron) as well as the hadronic
model is well known. It should be stressed that the difference is mostly
in the normalization, not in the functional dependence on $DG$. 
This is shown clearly by the muon signals plotted relative to
proton-QGSJetII, \reffig{Smu-ratios}. 

%This will be instrumental in our determination of the muon signal
%normalization from the data.

%\begin{figure}[t]
%\begin{center}
%\noindent
%\includegraphics [width=0.45\textwidth]{icrc0752_fig03.eps}
%\end{center}
%\caption{Muon ground signals vs. distance to
%ground of the shower maximum, for proton (black) and iron (red) showers.}
%\label{fig:Smu-DG}
%\end{figure}

\section{Determining the muon normalization and energy scale}
\label{sec:cic}

The universality of the electromagnetic ground signal as well as of the
{\it evolution} of the muon signal can be used to parameterize the total ground
signal in a model- and primary-independent way. The signal at a fixed core
distance is then only a function of primary energy, distance to ground $DG$,
zenith angle $\theta$, and the overall muon normalization. The slight primary-dependence
of the electromagnetic signal enters as a systematic uncertainty in the method.
Given the measured {\it average} depth of shower maxmimum $\Xmaxavg$ as a 
function of energy, either from a fluorescence
detector (on site or a separate experiment) or from ground observables, the distance to ground can be directly determined 
from the zenith angle for each shower: $DG = X_0/\cos\theta - \Xmaxavg$,
where $X_0$ stands for the vertical depth of the atmosphere at the experiment
site. We parametrize the electromagnetic and muon signal as 
separate Gaisser-Hillas type functions in $DG$, leaving a normalization
factor free for the muon signal:
\begin{eqnarray}
S(E,\theta) &=& \Sem(E, \theta, \Xmaxavg) \nonumber\\
            &+& \Nmu(E)\cdot\Smu(\theta, \Xmaxavg)
\label{eq:Sparam}
\end{eqnarray}
Here, $\Smu$ denotes a reference muon signal, which we take to be 
proton-QGSJetII at 10$^{19}$~eV, and $\Nmu(E)$ is the relative muon
normalization at this energy. Hence, the energy $E$ and $\Nmu(E)$
are the remaining unknowns, which however cannot be disentangled for 
individual events in a ground array.

\begin{figure}[t]
\begin{center}
\noindent
\includegraphics [width=0.45\textwidth]{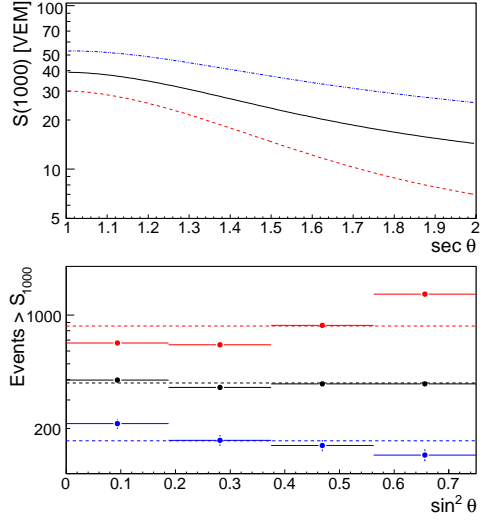}
\end{center}
\caption{{\it Upper panel:} the signal parametrization \refeq{Sparam} vs. 
$\sec\:\theta$ for different $\Nmu$ (black/solid$-$1.1, red/dashed$-$0.5, 
blue/dotted$-$2.0). {\it Lower panel:} 
histograms of number of events above the parametrized signal in equal
exposure bins, obtained for the same $\Nmu$ as shown in the upper panel
from a Monte Carlo data set (see text).}\label{fig:CIC}
\end{figure}

\reffig{CIC} (upper panel) shows the zenith angle dependence of the signal 
(\refeq{Sparam}) for a fixed energy of 10$^{19}$ eV and different values of 
$N_\mu$. It is evident that the smaller the $N_\mu$, the steeper the 
$\theta$ dependence is.
We can now use the fact that, within statistics, the arrival
directions of high energy cosmic rays are isotropic. Therefore, we 
divide the ground detector data set in equal exposure bins in zenith angle
($\sin^2 \theta$ bins). Given a muon normalization, we calculate the number 
of events in each bin above a given reference energy 
(here $E_{\rm ref}$=10$^{19}$~eV), using \refeq{Sparam}. We then adjust 
$\Nmu(E_{\rm ref})$ in the signal 
parametrization \refeq{Sparam} to that value which gives an equal number of
events $N(>S(E_{\rm ref}, \theta))$ in each zenith angle bin (lower panel in 
\reffig{CIC}). For a range of $\Nmu$ values, we calculate the $\chi^2$/dof 
of the event
histogram relative to a flat distribution in $\sin^2\theta$. This determines 
the experimental value of $N_\mu$ and its errors.
Once $N_\mu$ is determined, \refeq{Sparam} can be used
to set the energy scale of the experiment.

In order to prove the feasibility of this method, we have simulated
1,000 realizations of a ground array data set with $\sim$2,000 events above 
$10^{19}$ eV, distributed according to the observed cosmic ray spectrum and 
for different primary compositions (pure proton, iron, or mixed composition).
The zenith angle of each shower is sampled from a flat $\sin^2\theta$ 
distribution, while $\Xmax$ is obtained from the distributions predicted by 
QGSJetII for each primary and energy. $\Nmu$ is fluctuated according to the 
model predictions. Note that the magnitude of fluctuations in
$\Xmax$ and $\Nmu$ are only dependent on the primary particle, not the
hadronic model.
 
\refeq{Sparam} is then used to calculate the signal at 1000 m from the shower 
core, $S(1000)$, which is also smeared with an experimental reconstruction 
accuracy
(10\% for high signals, and increasing rapidly at signals less than 10 VEM).
We then applied the method described above to calculate the muon 
normalization for each simulated data set.
We found that $\Nmu$ is systematically slightly {\it overestimated}, with the 
bias mainly depending on composition,
and only weakly on the detector resolution. For pure proton composition,
the bias at $10^{19}$~eV was found to be around 14\% of the true $\Nmu$ value, 
while for pure iron, it only amounts to a few percent, due to the much smaller
fluctuations of iron showers. This bias can then
be subtracted from the determined $N_{\mu,\rm exp}$ to obtain an estimate
of the true $\Nmu$, the uncertainty in the bias entering as an additional
contribution to the systematic error. Note however that a knowledge of 
$\Xmaxavg$ already places strong constraints on the composition.

Taking into account this knowledge, and assuming the observed universality
violation and an error of $\Xmaxavg$ of $\sim15\:\gcmsq$, we found
that the total systematic uncertainty of $\Nmu$ achievable is less than 10\%,
roughly the statistical error of $\Nmu$ for this data set.

\section{Conclusions}

Assuming that air shower univerality holds, the method presented allows for a 
measurement of the muon content of air showers to better than 10\% for
existing experiments. With similar precision, it also determines a
converter of signal at ground to energy, i.e. a
model-independent ground detector energy scale. In addition, the
measurement of $\Nmu$ can be performed at any energy accessible to the
experiment. The measured evolution of $\Nmu(E)$ is a further observable of
relevance to hadronic models and composition.
This method has been applied to data from the Pierre Auger Observatory
 \cite{nmu}  yielding results that constrain hadronic interaction models.

%This is the reference to .bib file (Without .bib!)
\bibliography{icrc0752-arxiv}

\begin{thebibliography}{10}

\bibitem{Hillas}
{A. M. Hillas} et~al.
\newblock {\em Acta Physica Acad. Scient. Hung.}, 29, Suppl.~3:533, 1970.

\bibitem{Newton}
{D. Newton} et~al.
\newblock {\em Astrop. Phys.}, 26:414--419, 2007.

\bibitem{Chou}
A.~S. Chou et~al.
\newblock In {\em Proc. 29th ICRC, Pune}, volume~7, page 319, 2005.

\bibitem{Nerling}
{F. Nerling} et~al.
\newblock {\em Astrop. Phys.}, 24:421--437, 2006.

\bibitem{Giller}
{M. Giller} et~al.
\newblock {\em Int. J. Mod. Phys. A}, 20:6821--6824, 2005.

\bibitem{Gora}
{D. Gora} et~al.
\newblock {\em Astrop. Phys.}, 24:484--494, 2006.

\bibitem{corsika}
{D. Heck} et~al.
\newblock {\em Forschungszentrum Karlsruhe Report FZKA 6019}, 1998.

\bibitem{qgsjet2}
S.~Ostapchenko.
\newblock {\em Phys. Rev. D}, 74:014026, 2006.

\bibitem{fluka}
{A. Fass\`o} et~al.
\newblock {\em Proc. Int. Conf. on Adv. Monte Carlo Radiation Phys. (MC2000),
  Lisbon, Portugal}, Oct 2000.

\bibitem{gheisha}
H.~Fesefeldt.
\newblock {\em Report PITHA-85/02, RWTH Aachen}, 1985.

\bibitem{sibyll}
{R. S. Fletcher} et~al.
\newblock {\em Phys. Rev. D}, 50:5710--5731, 1994.

\bibitem{sibyll2}
{R. Engel} et~al.
\newblock {\em International Cosmic Ray Conference, Salt Lake City}, 1:415,
  1999.

\bibitem{nmu}
R.~Engel [Pierre~Auger Collaboration].
\newblock In {\em these proceedings, \#605}, 2007.

\end{thebibliography}
%This in the bibtex style, is ok.
\bibliographystyle{unsrt}
%plain}
\end{document}